\documentclass[]{spie}  

\usepackage{amsmath,amsfonts,amssymb}
\usepackage{graphicx}
\usepackage[colorlinks=true, allcolors=blue]{hyperref}

\title{From Colors to Chemistry: A Combined Lenslet/Slicer IFS for Medium-Resolution Spectroscopy}

\author[a,b]{R.~Deno Stelter}
\author[a,b]{Andrew~J. Skemer}
\author[c]{Cyril Bourgenot}
\affil[a]{UC Observatories, 1156 High Street, Santa Cruz, CA 95064 USA}
\affil[b]{UC Santa Cruz, 1156 High Street, Santa Cruz, CA 95064 USA}
\affil[c]{Durham University, Stockton Rd, Durham DH1 3LE UK}

\authorinfo{Further author information: (Send correspondence to R.D.S)\\R.D.S.: E-mail: deno@ucolick.org, Telephone: 1 831 459 5780}

\begin{document} 
\maketitle

\begin{abstract}
We present the design and lab performance of a prototype lenslet-slicer hybrid integral field spectrograph (IFS), validating the concept for use in future instruments like SCALES/PSI-Red. 
By imaging extrasolar planets with IFS, it is possible to measure their chemical compositions, temperatures and masses. Many exoplanet-focused instruments use a lenslet IFS to make datacubes with spatial and spectral information used to extract spectral information of imaged exoplanets. 
Lenslet IFS architecture results in very short spectra and thus low spectral resolution. 
Slicer IFSs can obtain higher spectral resolution but at the cost of increased optical aberrations that propagate through the down-stream spectrograph and degrade the spatial information we can extract. 
We have designed a lenslet/slicer hybrid that combines the minimal aberrations of the lenslet IFS with the high spectral resolution of the slicer IFS.
The slicer output f/\# matches the lenslet f/\# requiring only additional gratings.  
\end{abstract}


\section{INTRODUCTION}
\label{sec:intro}  
\begin{figure}[htp]
    \centering
    \includegraphics[width=0.98\textwidth]{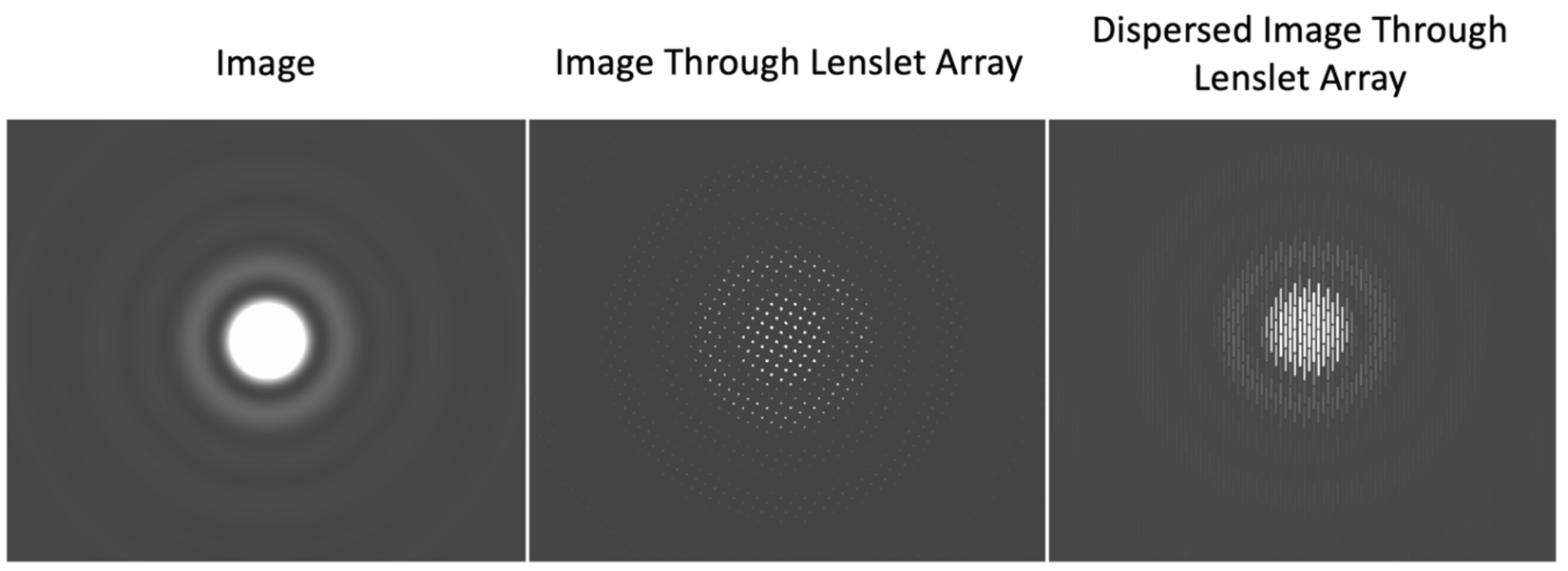}
    \caption{Example of how lenslet array IFS' work.
        At left, a white light diffraction-limited PSF.
        The center shows the same image sampled with a lenslet array, with the rows and columns rotated relative to the image axes by roughly 45 degrees.
        The dots are the lenslet pupil images, one per lenslet (the PSF is super-Nyquist sampled for demonstration purposes).
        The right panel shows the dispersed lenslet pupil images.
        Note that the spectra are short relative to the detector dimensions, and do not overlap.
  }
    \label{fig:low-resIFS}
\end{figure}

By imaging extrasolar planets around nearby bright stars, it is possible to measure the chemical compositions, temperatures and masses of these planets~\cite{barman2011,marley2012}.  
Most exoplanet-customized instruments use lenslet-based integral-field spectrographs (e.g., GPI~\cite{macintosh2014gpi}, SPHERE~\cite{beuzit2008sphere}, and CHARIS~\cite{groff2015charis}) to create images as a function of wavelength that can be used to construct a spectrum of the imaged planet.  
While lenslet-based spectrographs have already led to many exciting results, their architecture results in very short spectra on their detectors, and thus low spectral resolution, as shown in Figure~\ref{fig:low-resIFS}.  
Slicer integral field spectrographs can obtain higher spectral resolution and thus extract more information about the imaged planet, but at the cost of increased aberrations that propagate through the down-stream spectrograph and degrade the spatial information we can extract.  

\begin{figure}[htp]
    \centering
    \includegraphics[width=0.98\textwidth]{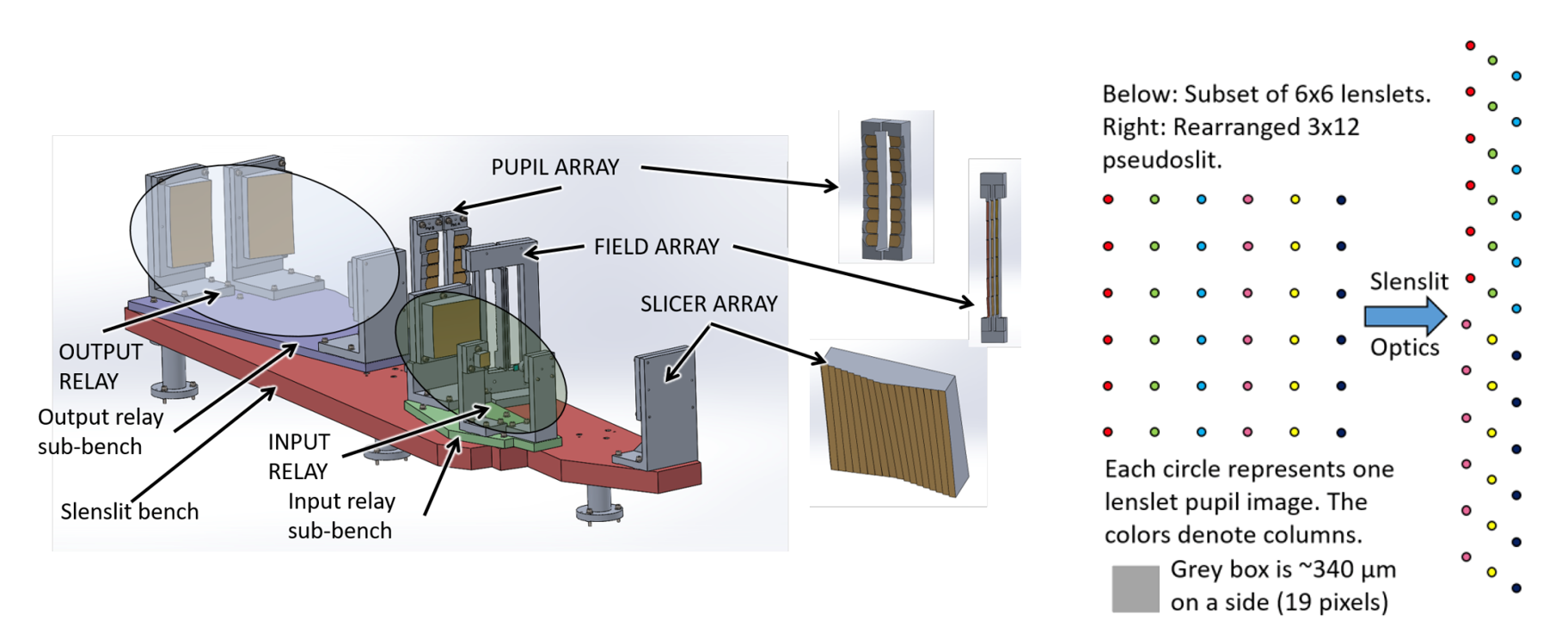}
    \caption{
        Left: Annotated SCALES slenslit opto-mechanical design with input and output relays circled, and the slicing optics shown.
        Right: A schematic of the increased packing factor using a 6x6 subset of lenslet arrays.
        For SCALES, the lenslet pitch is $\sim340~\mu$m (approximately 19 detector pixels), indicated by the grey box.
        The slenslit optics simply repackage the lenslet array pupil images into an interleaved 3-column pseudoslit with one-third the original spacing in the vertical direction.
    }
    \label{fig:slenslitPacking}
\end{figure}

\begin{figure}[htp]
    \centering
    \includegraphics[width=0.9\textwidth]{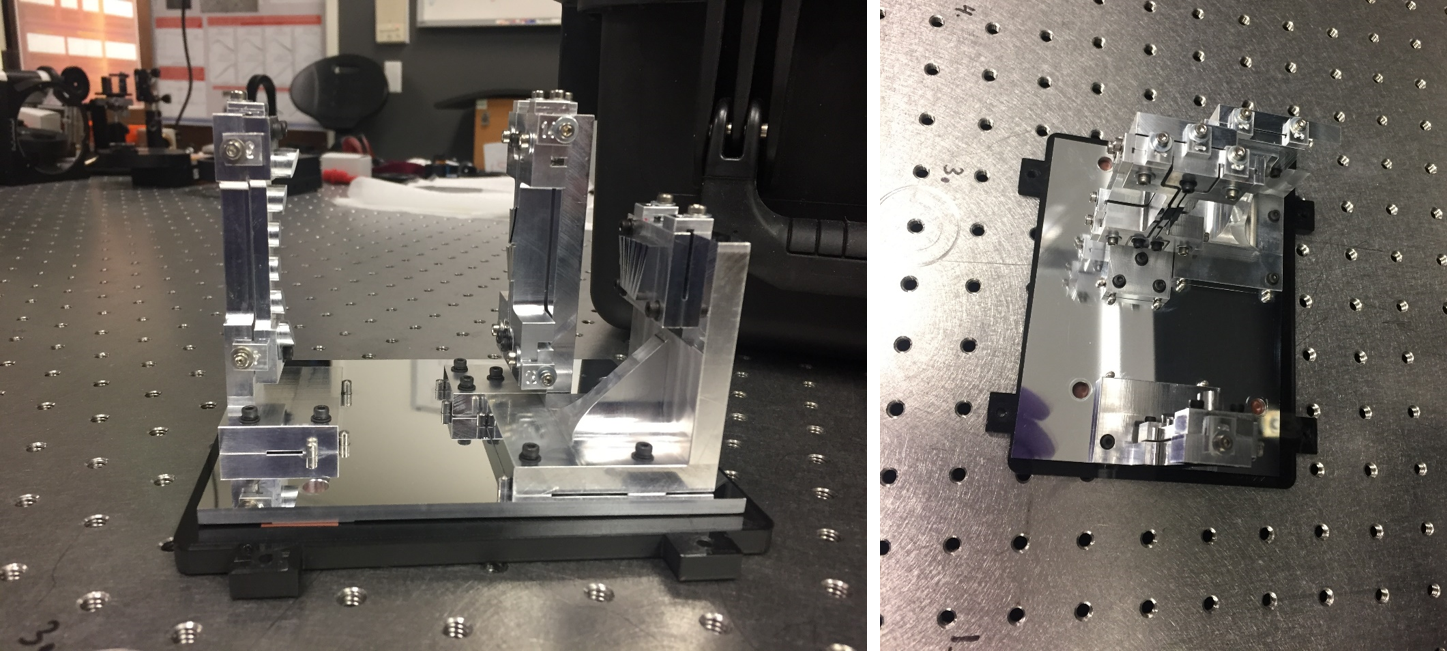}
    \caption{Slenslit prototype slicing optics on an optical table at the LAO.}
    \label{fig:slenslit-intro}
\end{figure}

\begin{figure}[htp]
    \centering
    \includegraphics[width=0.98\textwidth]
    {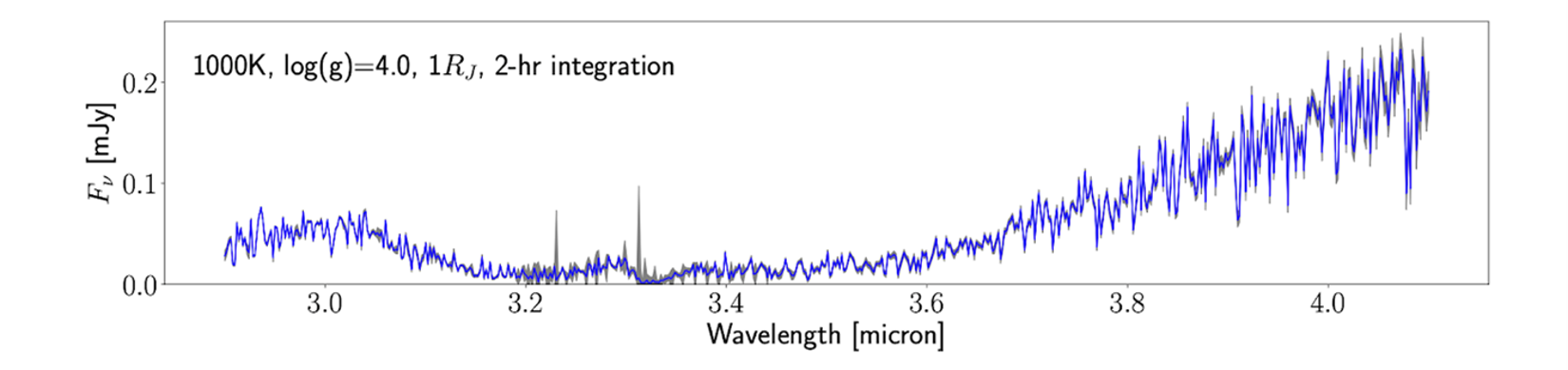}
    \caption{
        Simulation\cite{briesemeister2020} of mid-resolution spectroscopy of the exoplanet HR 8799 c, with the model atmosphere parameters for the exoplanet and exposure time listed.
        The signal is in blue with error bars in grey for each resolution element.
    }
    \label{fig:mid-resSpectro}
\end{figure}

We have designed a lenslet/slicer hybrid, dubbed a slenslit (Sliced LENslet pseudoSLIT), that combines the high optical quality of the lenslet-based approach with the high spectral resolution of the slicer approach for the Santa Cruz Array of Lenslets for Exoplanet Spectroscopy (SCALES) instrument\cite{stelter2018,briesemeister2020,stelter2020,sallum2021lpi}.
SCALES will operate behind the W.~M.~Keck Observatory's Adaptive Optics (AO) system over a bandpass of $2-5~\mu$m, and has a field of view of $2$ arcsec in diameter.
Figure~\ref{fig:slenslitPacking} shows the SCALES slenslit optics and a schematic of the increased packing density of lenslet pupil images.
The  SCALES slenslit optics  are insertable, allowing flexibility in trading field-of-view for spectral resolution.
In order to demonstrate the slenslit design is functional, we designed and fabricated a smaller benchtop prototype slenslit, shown in Figure~\ref{fig:slenslit-intro}.
Moving from low-resolution (R $\sim100$, shown in Figure~) to mid-resolution (R $\sim10\;000$) spectroscopy takes exoplanet characterization from the realm of distinguishing color to performing atmospheric chemistry. 
\begin{figure}[htp]
    \centering
    \includegraphics[width=0.7\textwidth]
    {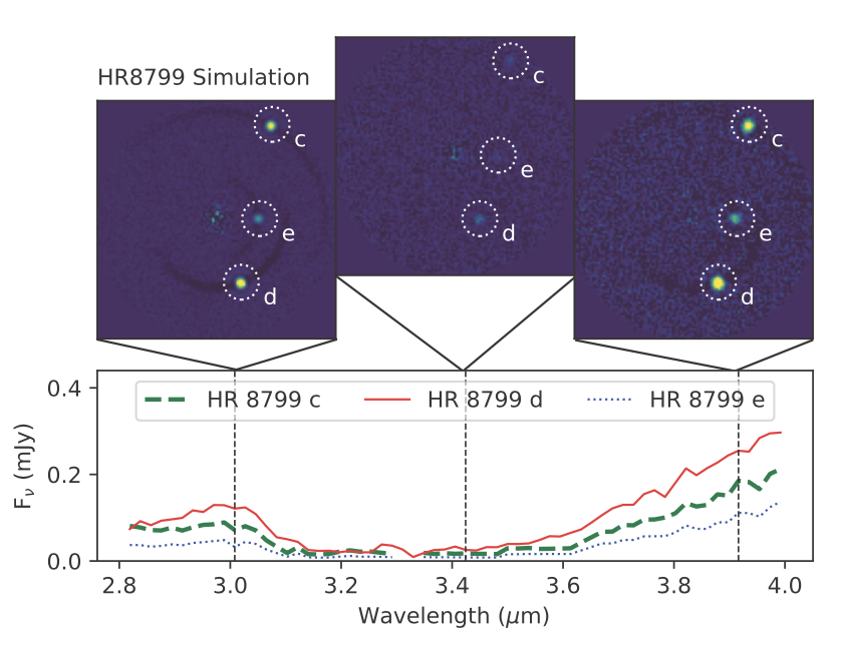}
    \caption{
        Simulation\cite{briesemeister2020} of low-resolution spectroscopy of the HR 8799 planetary system as imaged by SCALES.
        The top 3 images are extracted snapshots at specific wavelengths. 
        SCALES, like all IFS's, takes an spectrum at every point on the two-dimensional field of view; monochromatic images can be extracted from the datacube in post-processing.
        The plot below shows the spectra of the 3 innermost planets ($c$, $d$, and $e$) and the extracted image at 3 wavelengths.
    }
    \label{fig:low-resSpectro}
\end{figure}
Figures~\ref{fig:low-resSpectro} and \ref{fig:mid-resSpectro} illustrate the gain of information made accessible with the use of the slenslit.

This technology would be useful for future NASA-led space-based exoplanet imaging missions that require low spectral resolution for faint exoplanets but would benefit from higher spectral resolution on a handful of brighter targets.
We will describe the concept, design, initial lab performance, and next steps of the slenslit prototype in the following sections.

\section{Slenslit Concept}
\label{sec:slenslit-concept}
The slenslit  concept came about when pondering how to improve the spectral resolution of the SCALES instrument.
The realization that if one could just rearrange the image of the lenslets before light reached the disperser to make a pseudoslit, SCALES could achieve much higher spectral resolution led us to consider placing a set of slicing optics after the lenslet array.

In the case of SCALES, we chose to improve the `packing factor,' or center-to-center separation of lenslet pupil images, by a factor of 3 because the spectra separation when using the low-resolution mode is roughly 3 times smaller than the lenslet pitch.
We reformat the 18 columns of lenslet images into 3 interleaved `super-columns' with 6 columns each using slicing optics; each column has 17 rows of lenslet images.
The pseudo-slit thus has 3 lenslet images over the same distance as the original lenslet-to-lenslet distance.
The lenslet pupil images are magnified by an input three-mirror anastigmat (TMA) relay, geometrically rearranged and partially de-magnified by the slicing optics, and then returned back to the output pseudo-slit by the output TMA relay.
The pseudoslit is also parfocal with the original lenslet pupil image plane, which allows us to use the same spectrograph optics as the low-resolution mode; the SCALES slenslit opto-mechanical design and a schematic of the geometric rearrangement are shown in Figure~\ref{fig:slenslitPacking}.

Unlike other image slicers, the slenslit approach has much looser edge requirements; typically slicer mirrors must have very low ($<\sim5\%$) vignetting due to edge effects such as fratricide (where the construction of one mirror causes the tool to carve into the edges of its neighboring mirror).
By comparison, the slenslit field of view is not contiguous; the slice `sees' a discontinuous array of points corresponding to its column of lenslets.
The field mirrors, similarly, are not as concerned with edge effects, although the tight packing of the super-columns does place tight requirements on the positioning of the field mirrors.

\section{Slenslit Prototype Design and Construction}
\label{sec:slenslit-prototype-design}
\begin{figure}[htp]
    \centering
    \includegraphics[width=0.9\textwidth]{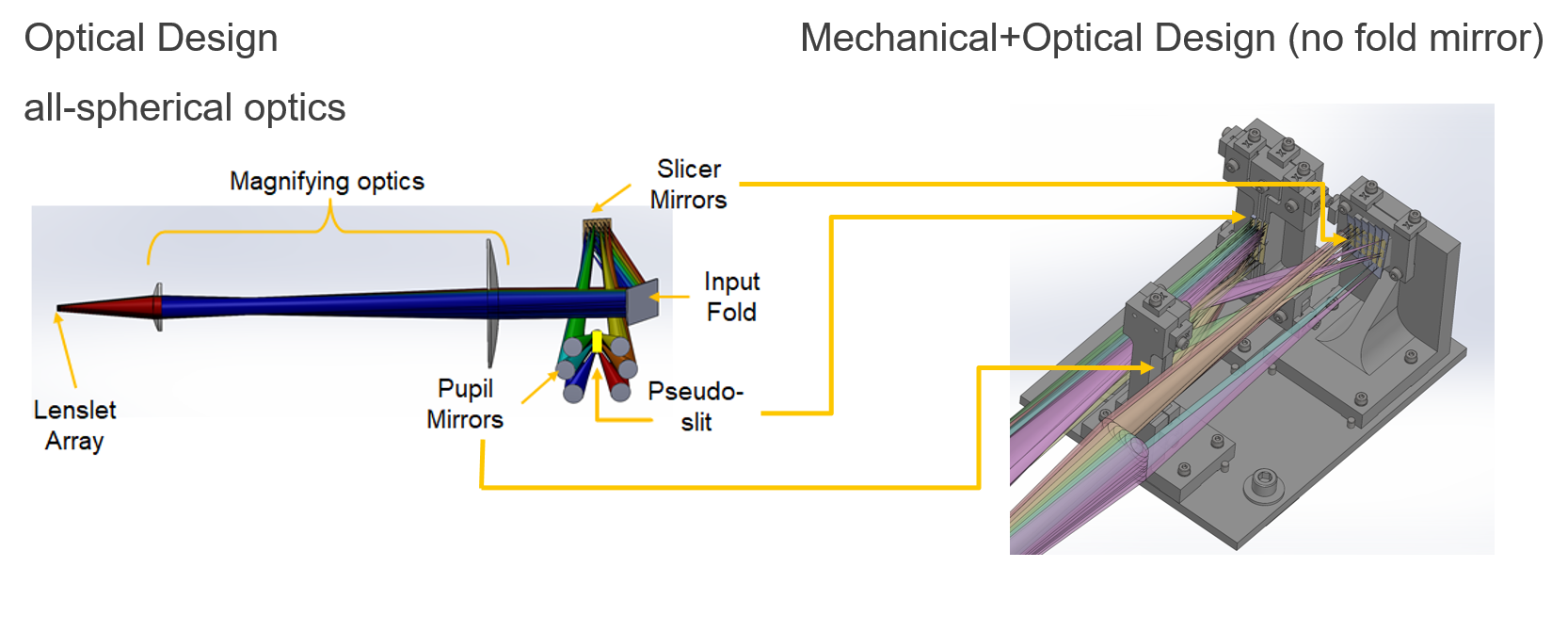}
    \caption{
    Left: Annotated optical design from Zemax.
    The slicing optics use all-spherical elements.
    Right: Annotated mechanical design (note that the optical beams, shown in pastel colors, did not always export properly).
    }
    \label{fig:slenslit-design}
\end{figure}

The slenslit prototype design was done in Zemax, followed by the mechanical design in SolidWorks.
The optical prescription was used to construct each mirror in SolidWorks using a monolithic construction (meaning the slicer, pupil, and field mirror arrays were cut into as few substrates as possible).
The optical beams were exported as STEP files and incorporated into the SolidWorks design, allowing us to verify the mechanical design matched the optical prescription.
Figure~\ref{fig:slenslit-design} shows the optical and mechanical designs.

The slicing optics are designed to work with an f/\#5.5, 1 mm pitch micro-lens array (or lenslet array) with square lenslets.
For this demonstrator, we chose to use a 6x6 subarray of lenslets, as this demonstrates the factor of 3 interleaving quite nicely.
Magnifying optics reimage the lenslet pupil images onto the slicer with a factor of 2 in magnification; slicers tend to work better with slower beam speeds.
The output of the slicing optics is then reimaged onto a screen or detector; the light can be dispersed with a prism or grating as desired.
The slices have identical radii of curvature, while the radii of curvature for the remaining mirrors was allowed to vary.

\begin{figure}[htp]
    \centering
    \includegraphics[width=0.95\textwidth]{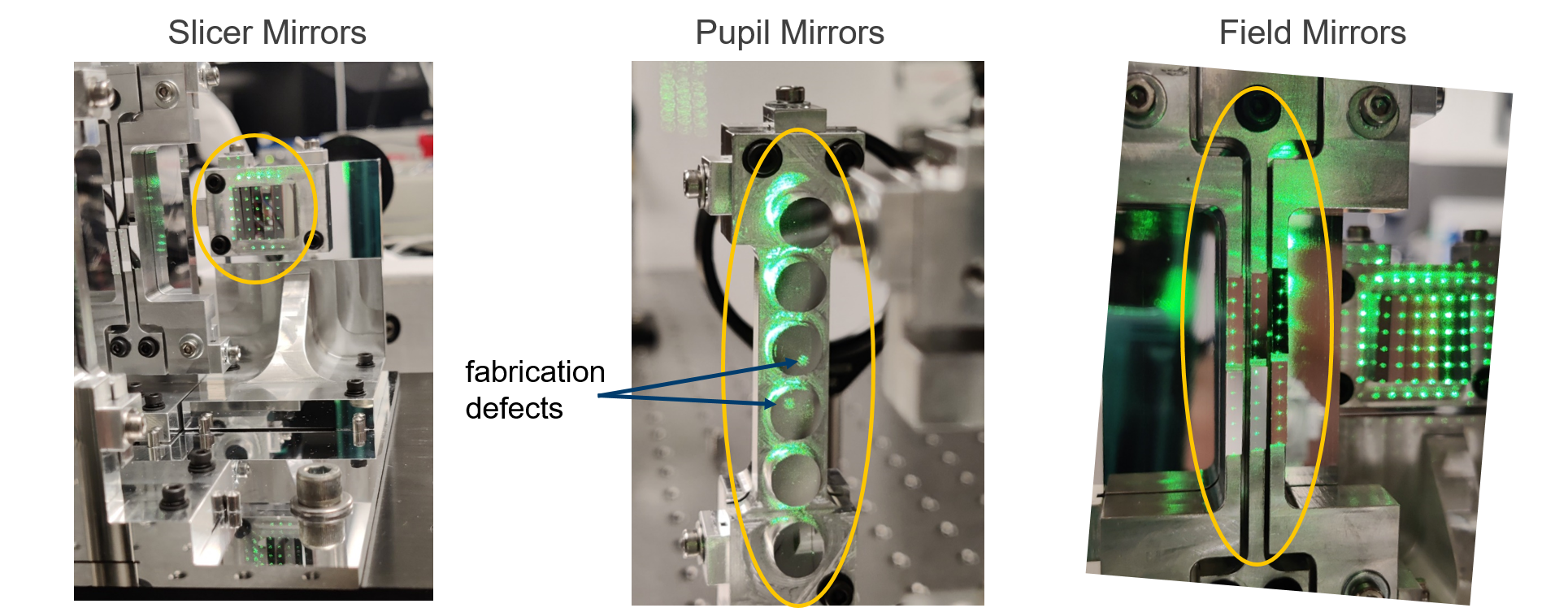}
    \caption{
        The slicing optics, illuminated with laser light through a 10x10 patch of lenslets.
        The two patches identified as fabrication defects are obvious areas of high scatter and discussed in the text.
    }
    \label{fig:mirrorDetails}
\end{figure}

We iterated on the mechanical design based on advice from Dr. Cyril Bourgenot at Durham Precision Optic (DPO), who fabricated the optics.
The alignment scheme uses a `bolt-and-go' approach, and builds on heritage of previous image slicers fabricated at DPO.
The mirrors are shown in Figure~\ref{fig:mirrorDetails}.
The bolt-and-go approach is a semi-kinematic mounting scheme that uses precision ground dowel pins and carefully machined mounting pads.
The mounting pads together define the plane of contact; using three planar features to define a plane over-constrains the plane, but is relatively straightforward to machine.
Likewise, the pins provide 3 line contacts; 2 line contacts one one side and a third on another.
This slightly over-constrains the position of the bracket, but emulates the minimum number of points needed to define a position on a plane.

Similarly, the use of 'n' shaped precision shims define the positions of the mirror array substrates on the mount, while providing conveniently measurable reference surfaces suitable for metrology and alignment.
Each bracket can be reinstalled on the bench, and each mirror substrate on its bracket, without disturbing the alignment at the $\sim$micron level.

Fabrication went smoothly, and with two exceptions, the surface roughness RMS is $\sim 5-15$ nm, with excellent figure error RMS (typically $< 0.1\lambda$); a few spots on some mirrors were higher, but given the prototype nature of this project were deemed to be acceptable.
Briefly, two of the pupil mirrors have fabrication defects resulting in higher surface roughness RMS, the origin of which is due to the fabrication technique, high speed diamond milling.
The fabrication defects are places where the local gradient of the mirror is nearly 0 with respect to the diamond tool axis normal, and due to the shape of the diamond tool and its interactions with a `flat' spot of the mirror, there is some deterioration of the surface roughness.
Other mirrors are not affected because the local gradients do not have a near-0 value.

\section{Slenslit Prototype Performance}
\label{sec:slenslt-performance}
\begin{figure}
    \centering
    \includegraphics[width=0.95\textwidth]{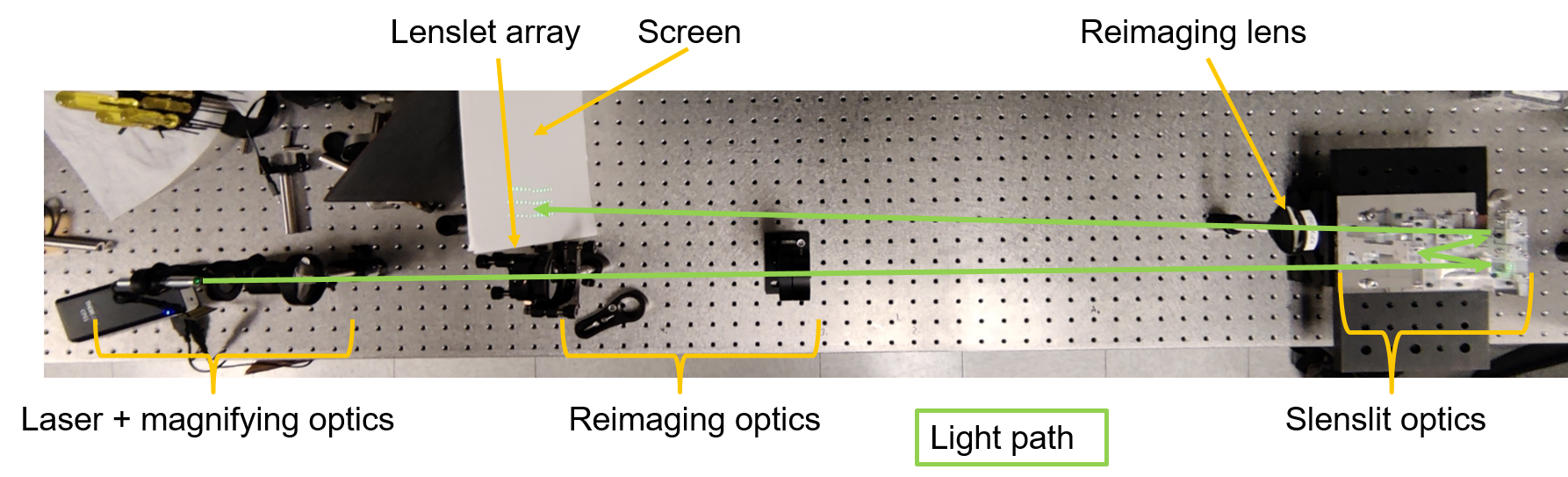}
    \caption{
        Experimental setup for the slenslit prototype.
    }
    \label{fig:setup}
\end{figure}

\begin{figure}
    \centering
    \includegraphics[width=0.9\textwidth]{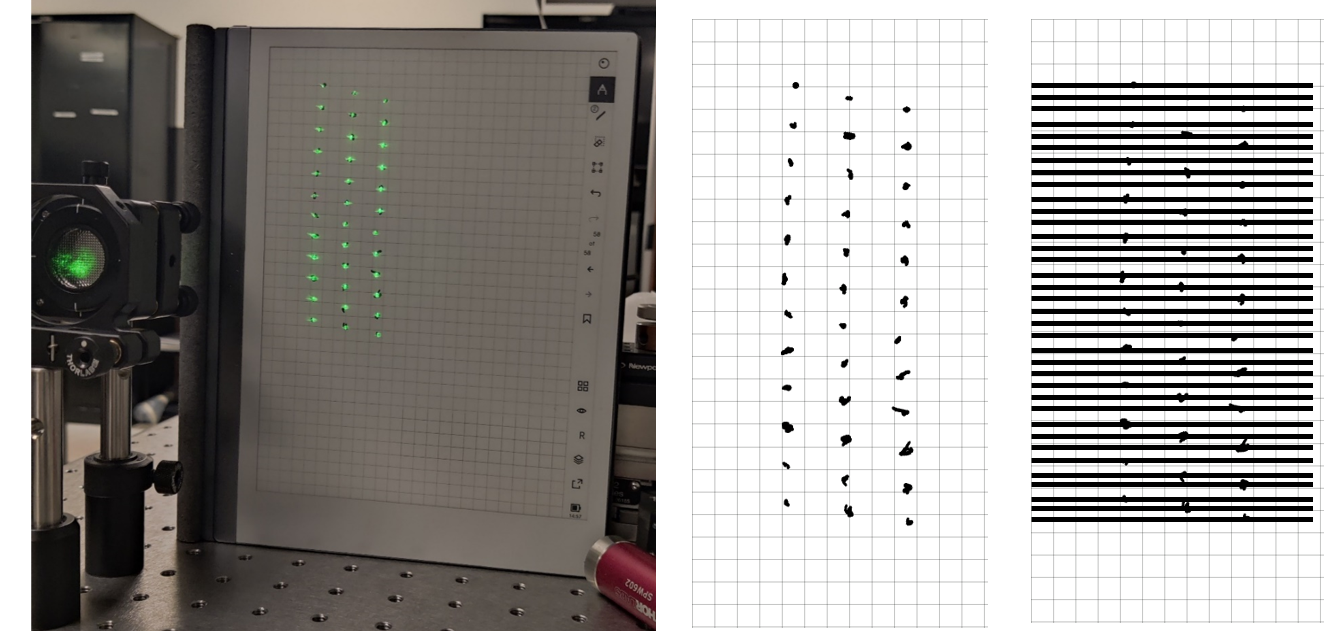}
    \caption{
        Left: Slenslit output as seen on a tablet screen.
        Note the light passing through the lenslet array is visible on the left in green light.
        Middle: We recorded the location of the lenslit pupil images on the tablet and digitized them.
        It is evident that the super-columns are slightly offset in the vertical direction.
        Right: As a quick check, drawing parallel lines through each of the lenslet pupil images to represent dispersed light shows that the spectra does not overlap.
        We note that the first author is not much of an artist.
    }
    \label{fig:output}
\end{figure}

We set the slenslit optics up in the Laboratory for Adaptive Optics (LAO) once the optics were received.
Figure~\ref{fig:setup} shows the setup on the optical table.
The experimental setup uses a green laser ($\lambda=532$ nm) which is magnified to fully illuminate the 6x6 subarray of lenslets used by the slenslit prototype.
The lenslets are reimaged with a zoom lens, which allowed us to adjust the plate scale during alignment of the slenslit.
After passing through the slicing optics, the slenslit output is reimaged onto a screen.
We used a reMarkable2 tablet as a screen to record the locations of the lenslet pupil images at the output, which let us verify the interleaving worked as desired.
Figure~\ref{fig:output} shows the recorded output and demonstrates that the interleaving is successful.

\section{Conclusion and Next Steps}
\label{sec:conclusion}
We have identified a suitable white light source and disperser for the slenslit prototype, which will allow us to directly test whether or not there is spectral overlap over the full optical bandpass; chromatic aberrations from the lenslet array, as well as the magnifying and reimaging optics, will limit the useful instantaneous bandpass.
Additionally we plan on placing the slenslit prototype on the LAO's SEAL test bed, which will allow us to inject simulated exoplanetary systems into the slenslit.
The SEAL test bed is a high-contrast AO instrument test bed and will require some pre- and post-optics to fit the slicing optics into the optical beam.
By using coronagraphs and other advanced exoplanet instrumentation tools, we can measure the actual spectral resolution improvement as a function of bandpass, contrast ratio, Strehl ratio, etc.

\acknowledgments 
 We gratefully acknowledge the support of the Mt. Cuba Astronomical Foundation which funded this project.
 Further, we would like to acknowledge the support of Dr. Phil Hinz, the Director of LAO, and the LAO staff scientists who graciously made room and offered advice on the experimental setup.
\bibliography{report} 

\begin{thebibliography}{1}

\bibitem{barman2011}
T.~S. {Barman}, B.~{Macintosh}, Q.~M. {Konopacky}, and C.~{Marois}, ``{Clouds
  and Chemistry in the Atmosphere of Extrasolar Planet HR8799b},'' {\em
  Astrophysical Journal}~{\bf 733}, p.~65, May 2011.

\bibitem{marley2012}
M.~S. {Marley}, D.~{Saumon}, M.~{Cushing}, A.~S. {Ackerman}, J.~J. {Fortney},
  and R.~{Freedman}, ``{Masses, Radii, and Cloud Properties of the HR 8799
  Planets},'' {\em Astrophysical Journal}~{\bf 754}, p.~135, Aug. 2012.

\bibitem{macintosh2014gpi}
B.~A. {Macintosh}, A.~{Anthony}, J.~{Atwood}, B.~{Bauman}, A.~{Cardwell},
  K.~{Caputa}, J.~{Chilcote}, R.~J. {De Rosa}, D.~{Dillon}, R.~{Doyon},
  J.~{Dunn}, D.~{Erickson}, M.~P. {Fitzgerald}, D.~T. {Gavel}, R.~{Galvez},
  S.~{Goodsell}, J.~{Graham}, A.~Z. {Greenbaum}, M.~{Hartung}, P.~{Hibon},
  P.~{Ingraham}, D.~{Kerley}, Q.~{Konopacky}, K.~{Labrie}, J.~{Larkin},
  J.~{Maire}, F.~{Marchis}, C.~{Marois}, M.~{Millar-Blanchaer}, K.~{Morzinski},
  A.~{Nunez}, R.~{Oppenheimer}, D.~{Palmer}, J.~{Pazder}, M.~{Perrin}, L.~A.
  {Poyneer}, L.~{Pueyo}, C.~{Quiroz}, F.~{Rantakyro}, V.~{Reshetov},
  L.~{Saddlemyer}, N.~{Sadakuni}, D.~{Savransky}, A.~{Serio},
  A.~{Sivaramakrishnan}, M.~{Smith}, R.~{Soummer}, S.~{Thomas}, J.~K.
  {Wallace}, J.~{Wang}, J.~{Weiss}, S.~{Wiktorowicz}, and S.~G. {Wolff}, ``{The
  Gemini planet imager: first light and commissioning},'' {\em Proceedings of
  the National Academy of Sciences} {\bf 111}, pp.~12661--12666, 2014.

\bibitem{beuzit2008sphere}
J.-L. {Beuzit}, M.~{Feldt}, K.~{Dohlen}, D.~{Mouillet}, P.~{Puget}, F.~{Wildi},
  L.~{Abe}, J.~{Antichi}, A.~{Baruffolo}, P.~{Baudoz}, A.~{Boccaletti},
  M.~{Carbillet}, J.~{Charton}, R.~{Claudi}, M.~{Downing}, C.~{Fabron},
  P.~{Feautrier}, E.~{Fedrigo}, T.~{Fusco}, J.-L. {Gach}, R.~{Gratton},
  T.~{Henning}, N.~{Hubin}, F.~{Joos}, M.~{Kasper}, M.~{Langlois}, R.~{Lenzen},
  C.~{Moutou}, A.~{Pavlov}, C.~{Petit}, J.~{Pragt}, P.~{Rabou}, F.~{Rigal},
  R.~{Roelfsema}, G.~{Rousset}, M.~{Saisse}, H.-M. {Schmid}, E.~{Stadler},
  C.~{Thalmann}, M.~{Turatto}, S.~{Udry}, F.~{Vakili}, and R.~{Waters},
  ``{SPHERE: a 'Planet Finder' instrument for the VLT},'' in {\em Ground-based
  and Airborne Instrumentation for Astronomy II},  I.~S. {McLean} and M.~M.
  {Casali}, eds., {\em Society of Photo-Optical Instrumentation Engineers
  (SPIE) Conference Series} {\bf 7014}, p.~701418, July 2008.

\bibitem{groff2015charis}
T.~D. {Groff}, N.~J. {Kasdin}, M.~A. {Limbach}, M.~{Galvin}, M.~A. {Carr},
  G.~{Knapp}, T.~{Brandt}, C.~{Loomis}, N.~{Jarosik}, K.~{Mede}, M.~W.
  {McElwain}, D.~B. {Leviton}, K.~H. {Miller}, M.~A. {Quijada}, O.~{Guyon},
  N.~{Jovanovic}, N.~{Takato}, and M.~{Hayashi}, ``{The CHARIS IFS for high
  contrast imaging at Subaru},'' in {\em Techniques and Instrumentation for
  Detection of Exoplanets VII},  S.~{Shaklan}, ed., {\em Society of
  Photo-Optical Instrumentation Engineers (SPIE) Conference Series} {\bf 9605},
  p.~96051C, Sept. 2015.

\bibitem{stelter2018}
R.~D. {Stelter}, A.~{Skemer}, and R.~{Kupke}, ``{Overview of the
  opto-mechanical design of the 2-5 micron arm of the Thirty Meter Telescope
  planetary systems imager},'' in {\em Ground-based and Airborne
  Instrumentation for Astronomy VII},  C.~J. {Evans}, L.~{Simard}, and
  H.~{Takami}, eds., {\em Society of Photo-Optical Instrumentation Engineers
  (SPIE) Conference Series} {\bf 10702}, p.~107029X, July 2018.

\bibitem{briesemeister2020}
Z.~{Briesemeister}, S.~{Sallum}, A.~{Skemer}, R.~D. {Stelter}, P.~{Hinz}, and
  T.~{Brandt}, ``{End-to-end simulation of the SCALES integral field
  spectrograph},'' in {\em Society of Photo-Optical Instrumentation Engineers
  (SPIE) Conference Series},  {\em Society of Photo-Optical Instrumentation
  Engineers (SPIE) Conference Series} {\bf 11447}, p.~114474Z, Dec. 2020.

\bibitem{stelter2020}
R.~D. {Stelter}, A.~J. {Skemer}, S.~{Sallum}, R.~{Kupke}, P.~{Hinz},
  D.~{Mawet}, R.~{Jensen-Clem}, C.~{Ratliffe}, N.~{MacDonald}, W.~{Deich},
  G.~{Kruglikov}, M.~{Kassis}, J.~{Lyke}, Z.~{Briesemeister}, B.~{Miles},
  B.~{Gerard}, M.~{Fitzgerald}, T.~{Brandt}, and C.~{Marois}, ``{Update on the
  preliminary design of SCALES: the Santa Cruz Array of Lenslets for Exoplanet
  Spectroscopy},'' in {\em Society of Photo-Optical Instrumentation Engineers
  (SPIE) Conference Series},  {\em Society of Photo-Optical Instrumentation
  Engineers (SPIE) Conference Series} {\bf 11447}, p.~1144764, Dec. 2020.

\bibitem{sallum2021lpi}
S.~{Sallum}, A.~{Skemer}, D.~{Stelter}, Z.~{Briesemeister}, N.~{Batalha},
  N.~{Batalha}, G.~{Blake}, T.~{Brandt}, W.~{Deich}, K.~{de Kleer}, I.~{de
  Pater}, J.~{Eisner}, M.~{Fitzgerald}, W.~{Fong}, B.~{Gerard},
  T.~{Greathouse}, T.~{Greene}, P.~{Hinz}, M.~{Honda}, R.~{Jensen-Clem},
  M.~{Kassis}, C.~{Kilpatrick}, G.~{Kruglikov}, R.~{Kupke}, M.~{Liu},
  J.~{Lyke}, N.~{MacDonald}, C.~{Marois}, D.~{Mawet}, B.~{Miles}, C.~{Morley},
  D.~{Powell}, C.~{Ratliff}, K.~{Sandstrom}, P.~{Sheehan}, J.~{Spilker},
  J.~{Stone}, K.~{Wagner}, and Y.~{Zhou}, ``{SCALES: Instrument Overview and
  Expected Science Outcomes},'' in {\em Lunar and Planetary Science
  Conference},  {\em Lunar and Planetary Science Conference}, p.~2412, Mar.
  2021.

\end{thebibliography}
\bibliographystyle{spiebib} 

\end{document}